\documentclass{ws-p8-50x6-00}

\begin{document}

\title{Channeling of Charged Particles Through Periodically Bent Crystals:
on the Possibility of a Gamma Laser
\footnote{
contributed to the conference ``Fundamental and
Applied Aspects of Modern Physics'' in Lüderitz, Namibia, 2001
}
}

\author{A. V. Korol, W. Krause, A. V. Solov'yov and  W. Greiner}

\address{Institut f\"ur Theoretische Physik der Johann Wolfgang
Goethe-Universit\"at, 60054 Frankfurt am Main, Germany
\\E-mail: solovyov@th.physik.uni-frankfurt.de}

\maketitle

\abstracts{We discuss radiation generated by positrons channeling in a
crystalline undulator. The undulator is produced by periodically
bending a single crystal with an amplitude much larger than the
interplanar spacing. Different approaches for bending the crystal are
described and the restrictions on the parameters of the bending are
established. We present the results of numeric calculations of the
spectral distributions of the spontaneous emitted radiation and
estimate the conditions for stimulated emission. Our investigations
show that the proposed mechanism provides an efficient source for high
energy photons, which is worth to be studied experimentally.}

\section{Introduction}
\label{sec:intro}

We discuss a new mechanism, initially proposed in
\cite{Korol98,Korol99}, for the generation of high energy photons by
means of the planar channeling of ultra-relativistic positrons through
a periodically bent crystal. In this system there appears, in addition
to the well-known channeling radiation, an undulator type radiation
due to the periodic motion of the channeling positrons which follow
the bending of the crystallographic planes. The intensity and the
characteristic frequencies of this undulator radiation can be easily
varied by changing the positrons energy and the parameters of the
crystal bending.
\begin{figure}[t]
\epsfxsize=25pc 
\epsfbox{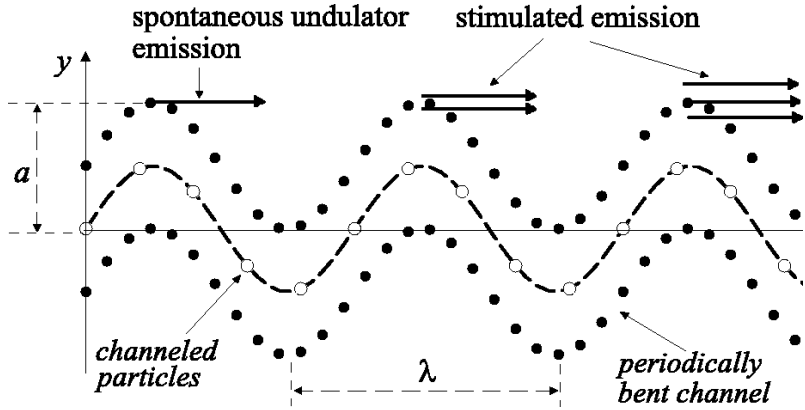} 
\caption{Schematic representation of spontaneous and stimulated
radiation due to positrons channeling in a periodically bent
crystal. The $y$- and $z$-scales are incompatible!
\label{fig:bent_crystal}}
\end{figure}

The mechanism of the photon emission by means of the crystalline
undulator is illustrated in Figure \ref{fig:bent_crystal}.  It is
important to stress that we consider the case when the amplitude $a$
of the bending is much larger than the interplanar spacing $d$ ($\sim
10^{-8}\;\mathrm{cm}$) of the crystal ($a \sim 10\ d$), and,
simultaneously, is much less than the period $\lambda$ of the bending
($a \sim 10^{-5} \dots 10^{-4}\, \lambda$).

In addition to the spontaneous photon emission by the crystalline
undulator, the scheme we propose leads to a possibility to generate
stimulated emission. This is due to the fact, that photons emitted at
the points of the maximum curvature of the trajectory travel almost
parallel to the beam and thus, stimulate the photon generation in the
vicinity of all successive maxima and minima of the trajectory.

\section{The bent crystal}
\label{sec:bent_crystal}

The bending of the crystal can be achieved either dynamically or
statically. In \cite{Korol98,Korol99} it was proposed to use a
transverse acoustic wave to dynamically bend the crystal. The
important feature of this scheme is that the time period of the
acoustic wave is much larger than the time of flight $\tau$ of a bunch
of positrons through the crystal. Then, on the time scale of $\tau$,
the shape of the crystal bending doesn't change, so that all particles
of the bunch channel inside the same undulator. One possibility to
couple the acoustic waves to the crystal is to place a piezo sample
atop the crystal and to use radio frequency to excite oscillations.

The usage of a statically and periodically bent crystal was discussed
initially in \cite{Korol98,Korol99} and later in \cite{Uggerhoj00}. In
the latter work the idea to construct a crystalline undulator based on
graded composition strained layers was suggested.
 
Let us now consider the conditions to be fulfiled in the stable
channeling regime.  The channeling process in a periodically bent
crystal takes place if the maximum centrifugal force in the channel,
$F_{\mathrm{cf}}\approx m \gamma c^2/R_{\mathrm{min}}$
($R_{\mathrm{min}}$ being the minimum curvature radius of the bent
channel), is less than the maximal force due to the interplanar field,
$F_{\mathrm{int}}$ which is equal to the maximum gradient of the
interplanar field (see \cite{Korol99}). More specifically, the ratio
$C=F_{\mathrm{cf}}/F_{\mathrm{int}}$ is better to keep smaller than
0.1, because otherwise the phase volume of channeling trajectories
becomes significantly reduced (see also \cite{Korol00}). The
inequality $C<0.1$ connects the energy of the particle, $\varepsilon=m
\gamma c^2$, the parameters of the bending (these enter through the
quantity $R_{\mathrm{min}}$), and the characteristics of the
crystallographic plane.
\begin{figure}[t]
\epsfxsize=25pc 
\epsfbox{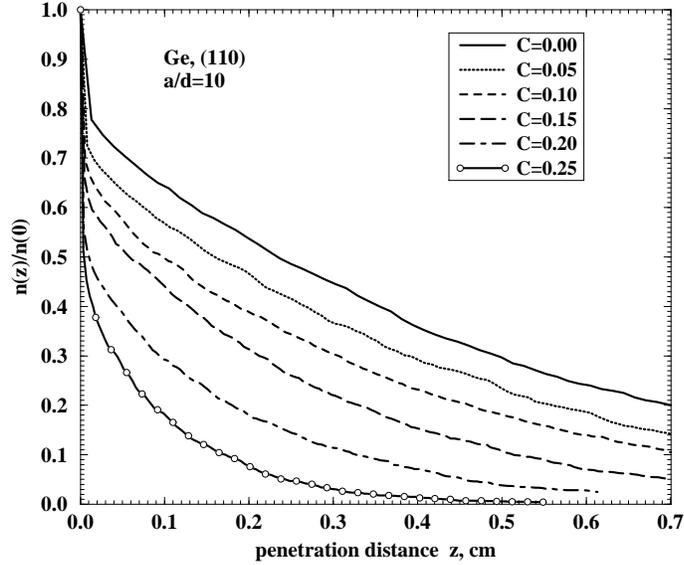} 
\caption{The calculated beam dencity 
dependences $n(z)/n(0)$ versus penetration distance $z$ for $5$ GeV
positrons channeling along the $(110)$ in Ge crystal for various
values of the parameter $C$ \protect\cite{Dechan01}.  The data
correspond to the shape function of the channel: $S(z)=a\sin(2\pi
z/\lambda)$.  The $a/d$ ratio equals $10$.  For each indicated $C$ the
corresponding values of $\lambda$, and the calculated magnitudes of
the dechanneling lengths $L_d^c$ and the number of undulator periods
$N_d^c=L_d^c/\lambda$ are presented in Table \protect\ref{Table1}.
\label{ld_ge}}
\end{figure}

A particle channeling in a crystal (straight or bent) undergoes
scattering by electrons and nuclei of the crystal. These random
collisions lead to a gradual increase of the particle energy
associated with the transverse oscillations in the channel. As a
result, the transverse energy at some distance $L_d$ from the entrance
point exceeds the depth of the interplanar potential well, and the
particle leaves the channel. The quantity $L_d$ is called the
dechanneling length \cite{Gemmel74}. To estimate $L_d$ one may follow
the method described in \cite{Biryukov96,Krause00a,Dechan01}. Thus, to
consider the undulator radiation formed in a crystalline undulator, it
is meaningful to assume that the crystal length does not exceed $L_d$.
The detailed numerical analysis of the dechanneling phenomena in
periodically bent crystals and its influence on the spectral
characteristics of the undulator radiation has been performed in
\cite{Dechan01}. An example of these calculations
for $5$ GeV positrons channeling along the $(110)$ in Ge crystal is
presented in figure \ref{ld_ge} and in \ref{Table1}.  In
\cite{Dechan01}, the similar calculations have been also performed for
the $Si$ and $W$ crystals.
\begin{table}

\begin{tabular}{@{}rrrrrrrrr}
Crystal:& Ge,   &$d=$  & $2.00$\AA,     &$R_c=$&$0.42$ cm &        &     &\\
  $C$   & $R_{min}$ &$\lambda$  &$L_d^e$ & $L_d^c$ &$N_d^e$&$N_d^c$&$\omega_1$&$p$\\
        &  cm     & $\mu$ m   &  cm    &  cm     &       &    &  MeV   & \\
 0.00   &$\infty$ &   -       &  0.263  & 0.513   &  -    &  -    &   -  &-\\
 0.05   & 8.465   & 81.8     &   0.237 & 0.450   &  29   &55  &1.37& 1.50\\ 
 0.10   & 4.232   & 57.8     &   0.213 & 0.364   &  36   &63  &1.26& 2.13\\
 0.15   & 2.822   & 47.2     &   0.190 & 0.269   &  40   &57  &1.15& 2.61\\
 0.20   & 2.116   & 40.9     &   0.168 & 0.176   &  41   &43  &1.05& 3.01\\
 0.25   & 1.693   & 36.6     &   0.148 & 0.095   &  40   &26  &0.98& 3.36\\
 0.30   & 1.411   & 33.4     &   0.129 & 0.060   &  38   &18  &0.92& 3.68\\
 0.35   & 1.209   & 30.9     &   0.111 & 0.028   &  35   & 9  &0.86& 3.98\\ 
 0.40   & 1.058   & 28.9     &   0.095 & 0.012   &  32   & 4  &0.82& 4.25\\
\end{tabular}
\caption{Dechanneling lengths for 5 GeV positron channeling
along the $(110)$ planes for the Ge crystal and various values of the
parameter $C$ \protect\cite{Dechan01}.  The data correspond to the
shape function $S(z)=a\sin(2\pi z/\lambda)$.  The $a/d$ ratio equals
$10$ except for the case $C=0$ (the straight channel).  The quantity
$L_d^c$ presents the accurately calculated dechanneling length
\protect\cite{Dechan01}, $N_d^c=L_d^c/\lambda$ is the corresponding
number of the undulator periods, $L_d^e$ is the dechanneling length
derived from a simple estimate (see
\protect\cite{Biryukov96,Krause00a,Dechan01}), $N_d^e=L_d^e/\lambda$.
Other parameters are: $d$ is the interplanar spacing, $R_c=
\varepsilon/ U_{\rm max}^{\prime}$ is the critical (minimal) radius
consistent with the condition $C \ll 1$, $\omega_1$ is the energy of
the first harmonic of the crystalline undulator radiation for the
forward emission, $p$ is the undulator parameter.  $R_{min}$ is the
minimum curvature radius of the bent channel centerline.  }
\label{Table1}
\end{table}

\begin{figure}[t]
\epsfxsize=25pc 
\epsfbox{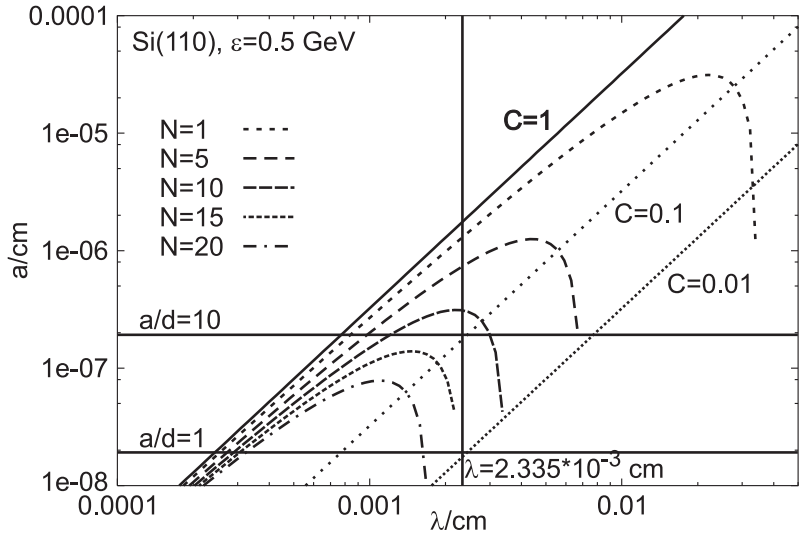} 
\caption{The range of parameters $a$ and $\lambda$ for a bent
Si(110) crystal at $\varepsilon=500$ MeV.
\label{fig:parameters}}
\end{figure}

Let us demonstrate how one can estimate, for a given crystal and
energy $\varepsilon$, the ranges of the parameters $a$ and $\lambda$
which are subject to the conditions formulated above.  For doing this
we assume that the shape of the centerline of the periodically bent
crystal is $a\,\sin(2\pi\,z/\lambda)$.  Figure \ref{fig:parameters}
illustrates the above mentioned restrictions in the case of
$\varepsilon=0.5$ GeV positrons planar channeling in Si along the
(110) crystallographic planes.  The diagonal straight lines correspond
to various values (as indicated) of the parameter $C$.  The curved
lines correspond to various values (as indicated) of the number of
undulator periods $N$ related to the dechanneling length $L_d$ through
$N=L_d/\lambda$.  The horizontal lines mark the values of the
amplitude equal to $d$ (with $d=1.92\cdot 10^{-8}\;\mathrm{cm}$ being
the (110) interplanar distance in Si) and to $10\,d$.  The vertical
line marks the value $\lambda = 2.335 \cdot 10^{-3}$ cm, for which the
spectra (see section \ref{sec:spectra}) were calculated.

\section{Spectra of the spontaneous emitted radiation}
\label{sec:spectra}

Let us consider the spectra of the spontaneous emitted radiation
calculated in \cite{Krause00a} using the quasiclassical method
\cite{Baier98}.  In \cite{Krause00a} the trajectories of the particles
were calculated numerically and then the spectra were evaluated.  The
latter include both radiation mechanisms, the undulator and the
channeling radiation.

\begin{figure}[t]
\epsfxsize=25pc 
\epsfbox{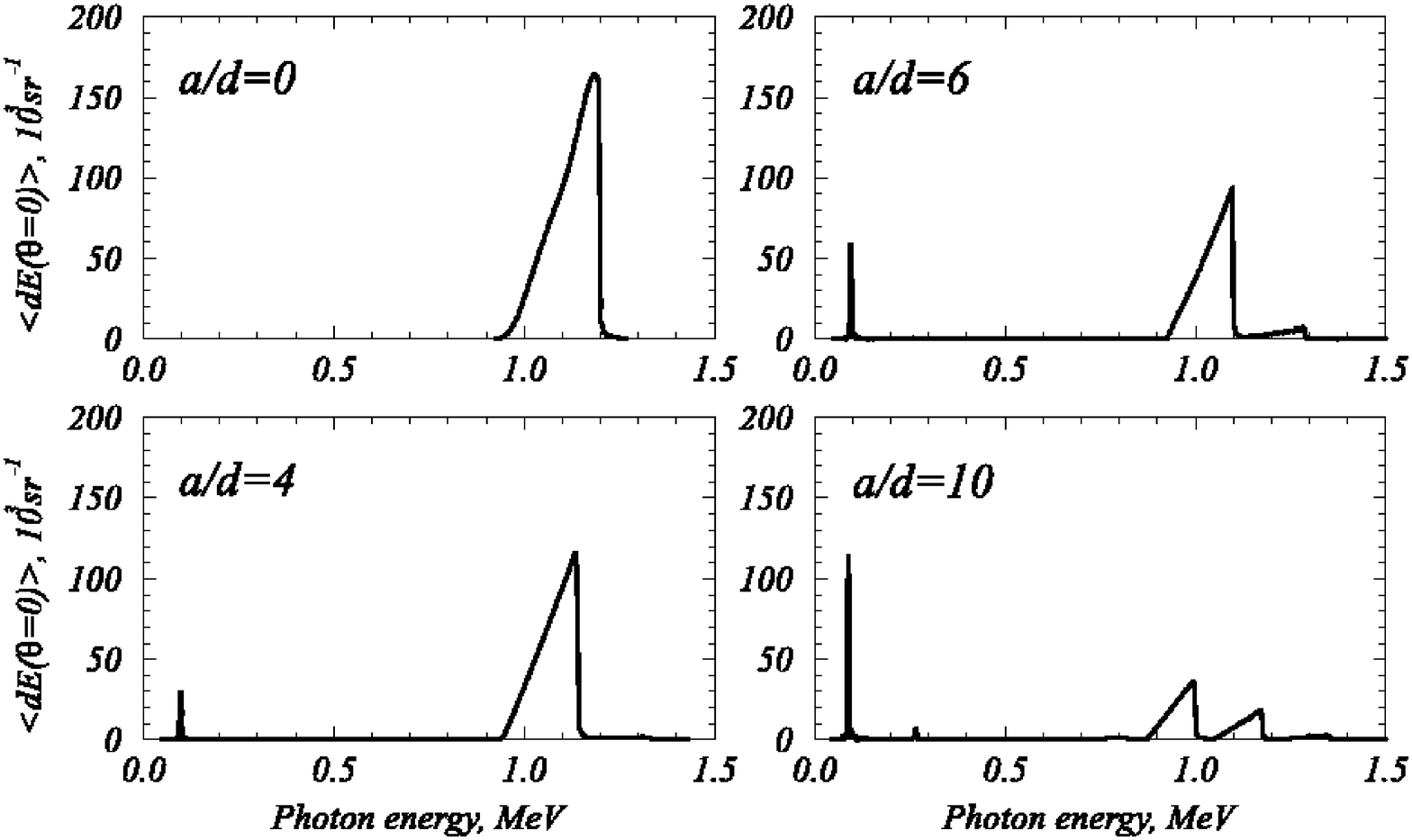} 
\caption{Spectral distributions of the total radiation emitted in
forward direction for $\varepsilon=500$ MeV positrons channeling in Si
along the (110) crystallographic planes for different $a/d$ ratios.
\label{fig:spectra}}
\end{figure}

The spectral distributions of the total radiation emitted in forward
direction for $\varepsilon=500$ MeV positrons channeling in Si along
the (110) crystallographic planes are plotted in figure
\ref{fig:spectra}. The wavelength is fixed at
$\lambda=2.335\cdot10^{-3}$ cm, while the ratio $a/d$ is changed from
0 to 10. The length of the crystal is $L_d=3.5\cdot10^{-2}$ cm and
corresponds to $N=15$ undulator periods.
The first graph in figure \ref{fig:spectra} corresponds to the case of
the straight channel ($a/d=0$) and, hence, presents the spectral
dependence of the ordinary channeling radiation only.  The spectrum
starts at $\hbar\omega\approx 960$ keV, reaches its maximum value at
$1190$ keV, and steeply cuts off at $1200$ keV.  This peak corresponds
to the radiation into the first harmonic of the ordinary channeling
radiation, and there is almost no radiation into higher harmonics.
The latter fact is consistent with general theory of dipole radiation
by ultra-relativistic particles undergoing quasiperiodic motion.  The
dipole approximation is valid provided the corresponding undulator
parameter $p_c = 2\pi \gamma(a_c/\lambda_c)$ is much less than 1.  In
this relation $a_c$ and $\lambda_c$ stand for the characteristic
scales of, correspondingly, the amplitude and the wavelength of the
quasiperiodic trajectory.  In the case of 0.5 GeV positrons channeled
along the (110) planes in Si one has $p_c\approx 0.2 \ll 1$ and all
the channeling radiation is concentrated within some interval in the
vicinity of the energy of the first harmonic.

Increasing the $a/d$ ratio leads to modifications in the spectrum of
radiation.  The changes which occur manifest themselves via three main
features, (i) the lowering of the ordinary channeling radiation peak,
(ii) the gradual increase of the intensity of undulator radiation due
to the crystal bending, (iii) the appearing of additional structure
(the sub-peaks) in the vicinity of the first harmonic of the ordinary
channeling radiation.  A more detailed analysis of these spectra can
be found in \cite{Krause00a}.

\section{Discussion of stimulated photon emission}
\label{sec:stimulated}

The scheme illustrated by figure \ref{fig:bent_crystal} allows to
discuss the possibility to generate stimulated emission of high energy
photons by means of a bunch of ultra-relativistic positrons moving in
a periodically bent channel. Indeed the photons emitted in the nearly
forward direction at some maximum or minimum point of the trajectory
by a group of particles of the bunch stimulate the emission of photons
with the same energy by another (succeeding) group of particles of the
same bunch when it reaches the next maximum/minimum.

In \cite{Korol99} estimates for the gain factor for the spontaneous
emission in crystalline undulators were obtained. It was demonstrated
that to achieve a total gain equal to 1 on the scale of the crystal
length (equal to the dechanneling length), one has to consider volume
densities $n$ of the channeling positrons on the level of
$10^{20}\dots 10^{21}$ cm$^{-3}$ for positron energies within the
range $0.5\dots 5$ GeV.  These magnitudes are high enough to be
questioned whether they can be really reached.

Let us estimate the volume density $n$ of a positron bunch which can
be achieved in modern colliders.  To do this we use the data presented
in \cite{PDG98} (see p.142) for a beam of 50 GeV positrons available
at SLC (SLAC, 1989).  The bunch length is 0.1 cm and the beam radius
is 1.5 $\mathrm{\mu m}$ (H) and 0.5 $\mathrm{\mu m}$ (V), resulting in
the volume of one bunch $V=2.4\cdot 10^{-9}\;\mathrm{cm^3}$.  The
number of particles per bunch is given as $4.0\cdot 10^{10}$.
Therefore, one obtains $n=1.7\cdot 10^{19}\;\mathrm{cm^{-3}}$.  This
value, although being lower by an order of magnitude than the
estimates obtained in \cite{Korol99}, shows that the necessary
densities should be reachable in the future with accelerators
optimized for high particle densities.

Finally let us discuss the required transverse emittance of the
beam. For doing this we need to consider the angle between the
particle's trajectory and the crystal plane. If this angle is larger
than the Lindhard angle $\Psi_{\mathrm{P}}$, the particle will not be
captured into the channeling mode and leaves the channel immediately
\cite{Lindhard}. For 5 GeV positrons channeling along the (110) plane
of silicon, we have $\Psi_{\mathrm{P}}=72\;\mathrm{\mu rad}$ and for
50 GeV positrons it equals to $\Psi_{\mathrm{P}}=23\;\mathrm{\mu
rad}$.

One can compare these values with the divergence of the SLC beam,
which transverse emittance in vertical direction is given as $0.05\,
\pi \;\mathrm{rad\ nm}$ \cite{PDG98}. With a vertical beam radius of
0.5 $\mathrm{\mu m}$ we get for the vertical beam divergence
$\Psi=100\;\mathrm{\mu rad}$. Thus the divergence of the beam is about
four times higher than the acceptance of the channel and so only a
quarter of all particles will participate in the channeling
process. Evidently it is necessary to reduce the divergence of the
beam, for example by increasing the beam radius. But then it is also
necessary to reduce the bunch length, to keep the particle density
high enough. Fortunately, like for the particle density, the values
achievable today differ only about one order of magnitude from the
values estimated above for the stimulated emission.

\section{Conclusion}
\label{sec:conclusion}

To conclude we point out that the crystalline undulators discussed in
this work can serve as a new efficient source for high energy
photons. As we have shown above, the present technology is nearly
sufficient to achieve the necessary conditions to constract not only
crystalline undulator, but also achieve the stimulated photon emission
regime.  The parameters of the crystalline undulator and the radiation
generated with its use differ substantially from what is possible to
achive with the undulators based on magnetic fields.

In our opinion the effects described above is worth experimental
study.  As a first step, one may concentrate on measurements of the
spontaneous undulator radiation spectra.

The related problems not discussed in this paper, but which are under
consideration, include the investigation of the crystal damage due to
the acoustic wave, photon flux and beam propagations.

\end{document}